\documentclass[11pt, onecolumn]{IEEEtran}

\usepackage{graphicx} % for pdf, bitmapped graphics files
\usepackage{amsmath} % assumes amsmath package installed
\usepackage{amssymb}  % assumes amsmath package installed
\usepackage{algorithm}
\usepackage{algorithmic}
\usepackage{cite}

\newtheorem{theorem}{Theorem}

\newtheorem{corollary}{Corollary}

\IEEEoverridecommandlockouts

\newcommand{\reals}{\mathbb{R}}

\newcommand{\expect}[1]{\mathbb{E}\left[#1\right]}

\newcommand{\prob}[1]{\mathbb{P}\left[#1\right]}

\newcommand{\pth}[1]{\left( #1 \right)}
\newcommand{\qth}[1]{\left[ #1 \right]}
\newcommand{\sth}[1]{\left\{ #1 \right\}}

\newcommand{\calB}{{\mathcal{B}}}

\newcommand{\calT}{{\mathcal{T}}}

\newcommand{\calX}{{\mathcal{X}}}

\newcommand{\eunorm}[1]{\left\|#1 \right\|}

\ifCLASSOPTIONcompsoc
\usepackage[caption=false,font=normalsize,labelfont=sf,textfont=sf]{subfig}
\else
\usepackage[caption=false,font=footnotesize]{subfig}
\fi

\begin{document}

\title{Trusted Multi-Party Computation and Verifiable Simulations: A Scalable Blockchain Approach}

\author{
    \IEEEauthorblockN{
    Ravi Kiran Raman,\IEEEauthorrefmark{1}\IEEEauthorrefmark{2}
    Roman Vaculin,\IEEEauthorrefmark{2} 
    Michael Hind,\IEEEauthorrefmark{2} 
    Sekou L. Remy,\IEEEauthorrefmark{2} 
    Eleftheria K. Pissadaki,\IEEEauthorrefmark{2}\\
    Nelson Kibichii Bore,\IEEEauthorrefmark{2} 
    Roozbeh Daneshvar,\IEEEauthorrefmark{2} 
    Biplav Srivastava,\IEEEauthorrefmark{2} and 
    Kush R. Varshney\IEEEauthorrefmark{2}
    }
    
    \IEEEauthorblockA{\IEEEauthorrefmark{1}University of Illinois at Urbana-Champaign} ~~~~
    \IEEEauthorblockA{\IEEEauthorrefmark{2}IBM Research}
}

\maketitle

%%%%%%%%%%%%%%%%%%%%%%%%%%%%%%%%%%%%%%%%%%%%%%%%%%%%%%%%%%%%%%%%%%%%%%%%%%%%%%%%

\begin{abstract}

Large-scale computational experiments, often running over weeks and over large datasets, are used extensively in fields such as epidemiology, meteorology, computational biology, and healthcare to understand phenomena, and design high-stakes policies affecting everyday health and economy. For instance, the OpenMalaria framework is a computationally-intensive simulation used by various non-governmental and governmental agencies to understand malarial disease spread and effectiveness of intervention strategies, and subsequently design healthcare policies. Given that such shared results form the basis of inferences drawn, technological solutions designed, and day-to-day policies drafted, it is essential that the computations are validated and trusted. In particular, in a multi-agent environment involving several independent computing agents, a notion of trust in results generated by peers is critical in facilitating transparency, accountability, and collaboration. Using a novel combination of distributed validation of atomic computation blocks and a blockchain-based immutable audits mechanism, this work proposes a universal framework for distributed trust in computations. In particular we address the scalaibility problem by reducing the storage and communication costs using a lossy compression scheme. This framework guarantees not only verifiability of final results, but also the validity of local computations, and its cost-benefit tradeoffs are studied using a synthetic example of training a neural network.

\end{abstract}

\maketitle
%%%%%%%%%%%%%%%%%%%%%%%%%%%%%%%%%%%%%%%%%%%%%%%%%%%%%%%%%%%%%%%%%%%%%%%%%%%%%%%%

\section{Introduction}

Machine learning, data science, and large-scale computations in general has created an era of computation-driven inference, applications, and policymaking \cite{Power2016, Shah2016}. Technological solutions, and policies with far-reaching consequences are increasingly being derived from computational frameworks and data. Multi-agent sociotechnical systems that are tasked with working collaboratively on such tasks function by interactively sharing data, models, and results of local computation.

However, when such agents are independent and lack trust, they might not collaborate with or trust the validity of reported computations of other agents. Quite often, these computations are also expensive and time consuming, and thus infeasible for recomputation by the doubting peer as a general course of action. In such systems, creating an environment of trust, accountability, and transparency in the local computations of individual agents promotes collaborative operation.

For instance, consider training a deep neural network with a given architecture using stochastic gradient descent (SGD). Here, the model and computations are deterministic given the data used for gradient computation. Applications are primarily interested in using the trained model represented by the weights of the trained network. But, if they lack trust in the training agent, they have no simpler way to verify the network than to retrain it. This is often impractical since the  (re)training process consumes extensive amounts of time and tends to require the use of specialized hardware like GPUs or TPUs. It is thus important to establish trust in the computations involved in the training phase.

To emphasize the importance of trust in multi-agent systems, let us also consider the case of policy design for malaria. OpenMalaria (OM) \cite{SmithMRPCSSGLTST2008} is an open source simulation environment, collaboratively developed to study malaria epidemiology and the effectiveness of control mechanisms. It is used extensively to design policies to tackle the disease. Here, individual agencies propose hypotheses regarding the disease and/or intervention policies, and study them by simulating them under specific environments \cite{PKSMKFSAHiette2016}. Considering the potential impact of such work in designing disease control policies, it is important to establish accountability and transparency in the process, so as to facilitate trusted adoption of results. Calls have been made for accountability and transparency in multi-agent computational systems, especially in high impact fields such as health \cite{Nelson2007}. A framework for decision provenance helps track the source of results, transparent computational trajectories, and a unified, trusted platform for information sharing. In fact, the US Centers for Disease Control \cite{CDC2005} states that:
\begin{quote}
public health and scientific advancement are best served when data are released to, or shared with, other public health agencies, academic researchers, and appropriate private researchers in an open, timely, and appropriate way. The interests of the public\dots transcends whatever claim scientists may believe they have to ownership of data acquired or generated using federal funds.
\end{quote}
This call implicitly assumes an inherent trust in the shared material. However, there exists significant disparity and inconsistency in current information-sharing mechanisms that not only hinder access, but also lead to questionable informational integrity \cite{VanPEGWHHB2014}. Here, trust and transparency are critical, but absent in current practice.

Establishing trust in computations translates to guaranteeing correctness of \textit{individual steps} of the simulation, and the integrity of the overall computational process leading to the reported results. Importantly, when computational models and parameters along with intermediate results of individual steps are shared, these steps can be validated by other agents who can recompute them, thereby validating the entire computation in a distributed manner.

Blockchain is a distributed ledger (database) technology that enables multiple distributed, potentially untrusted agents to transact and share data in a safe, secure, verifiable and trusted manner through mechanisms providing transparency, distributed validation, and cryptographic immutability \cite{CromanDEGJKMSSSSW2016}. As such, blockchain is the perfect tool for establishing the type of trust for complex, long running computations of interest. In this paper we use blockchain to record, share, and validate frequent audits (model parameters  with the intermediate results) of individual steps of the computation. We describe how blockchain-based distributed validation and  shared, immutable storage can be used and extended to enable efficient trusted verifiable computations. 

A common challenge arising in blockchain-based solutions is its scalability. The technology calls for significant peer-to-peer interaction and local storage of large records that include all the data generated in the network. These fundamental requirements result in significant communication and storage costs respectively. Thus, using the technology for large-scale computational processes over a large multi-agent networks is prohibitively expensive. In this paper, we address this scalability challenge by developing a novel compression schema and a distributed, parallelizable validation mechanism that is particularly suitable for such large-scale computation contexts.

\section{Prior Work}

We now provide a brief summary of prior work in related areas.

A variety of applications with widespread impact are being designed with the help of improved computational capabilities, easier access to data, and machine learning algorithms. Taking the context of malaria, as studied through OpenMalaria simulations, new pipelines for integrating AI tools and algorithms have been considered \cite{RemyBB2018}. Regression-based methods for better policy search have also been integrated with the open-source platform \cite{BentRRW2017}.

Considering the impact of such simulations, researchers have recently raised alarm over their lack of reproducability. Reproducing results from research papers in AI have been found to be challenging as a significant fraction of hyperparameters and model considerations are not documented \cite{GundersonK2018}. In another paper focused on reproduction of results in deep learning \cite{HendersonIBPPM2017}, the authors explore the possible reasons, and cite variability in evaluation metrics and reporting among different algorithms and implementations.

Accountability and transparency are being increasingly sought after in large-scale computational platforms, with particular focus on establishing tractable, consistent computational pipelines. The problem of establishing provenance in decision-making systems has been considered \cite{SinghCN2018} through the use of an audit mechanism. Distributed learning in a federated setting with security and privacy limitations has also been considered recently \cite{VermaCC2018}.

In fact, the problem of trust in multi-agent computational systems was considered at the beginning of the $20$th century from the viewpoint of reducing errors in complex calculations performed by human workers \cite{Grier2011}. Large-scale computational problems were solved using redundant evaluation of smaller sub-tasks assigned to human workers, and verified using computational checkpoints. We can draw significant insight into reliable distributed computing from these practices.

Blockchain systems have brought forth the means for creating distributed trust in peer-to-peer networks for transactional systems \cite{Androulaki2018}. A variety of applications that invoke interactions and transactions among untrusting agents have benefited from the trust enabled by blockchains \cite{TapscottT2016,IansitiL2017,Vukolic2017}. More recently, blockchains have been used in creating secure, distributed, data sharing pipelines in healthcare \cite{Tsai2018} and genomics \cite{OzercanIAA2018}. This trust can also be leveraged in creating trusted distributed computing systems, as highlighted in this paper.

\section{Motivation} \label{sec:motivation}
 
The motivation behind studying the problem of trust in multi-agent computational systems is best understood through the practical example of OpenMalaria. Agencies making policy decisions gain access to research findings such as transmission models from the work of independent organizations studying malaria around the world. The open source nature of the platform has facilitated widespread access, and has created a large-scale collaborative effort toward tackling the disease. The studies performed by various agents lead to policies that determine the health and well-being of vast sections of the community. Sharing data, models, and outcomes of simulation studies thus requires accountability and transparency with the guarantee of computational integrity, enabling the creation of reliable distributed computing platforms.

Let us consider a simple experiment that a malaria data scientist (MDS) is interested in conducting, to study the disease spread and control under a specific environment characterized by factors such as demographics, entomology, and intervention strategies in place. In particular, the MDS wishes to evaluate intervention strategies, such as distributing insecticide treated nets (ITN) and commissioning indoor residual spraying (IRS). The simulation is used to evaluate the efficiency of the policies, in terms of quantities such as disability adjusted life years (DALY) which quantifies the total life years lost from malaria-related fatality. Each policy also incurs a related cost of implementation, such as healthcare system costs (HSC) and intervention costs (IC). 

By studying the cost-benefit tradeoffs, the data scientist and/or other agencies can design optimal policies. Such agencies have access to simulation results performed by independent agents/workers. However, malicious agents, such as one who manufactures ITNs, could generate spurious results that address vested interests, rather than providing accurate insight into the disease. At the same time, workers with insufficient computational resources might also generate errors in the simulation process, potentially generating wrong inferences about the disease and policies.

Adoption of such results for policy design requires inter-agent trust, which is not guaranteed in multi-agent systems. Repeating experiments for each adoption is prohibitively expensive. Trust in computations would significantly assist information sharing. 

The notion of trust has been considered from a variety of standpoints \cite{FalconeST2001} and has contextually varied definitions as considered in depth in \cite{Marsh1994}. A qualitative definition of trust in multi-agent computational systems can be adapted from \cite{Dasgupta2000, RamchurnHJ2004} as:
\begin{quote}
\emph{Trust} is the belief an agent has that the other party will execute an agreed upon sequence of actions and reports an accurate representation of computed result (being honest and reliable).
\end{quote}
We provide a more specific characterization of trust. Such computations in general are composed of a sequence of atomic operations that update a system state iteratively. For instance, this could be OM simulations tracking the progression of malaria in a certain community, or the weights of a neural network as updated iteratively by a training algorithm. Establishing distributed trust, as defined, for such computations in a \emph{universal} sense (without contextual understanding of computation specifics) requires checking consistency of individual steps of the simulation by recomputation. In particular, we decompose trust into two main components:
\begin{itemize}
\item {\bf Validation:} The individual atomic computations of the simulation are guaranteed and accepted to be correct.
\item {\bf Verfication:} The integrity of the overall simulation process can be checked by other agents in the system.
\end{itemize}
The two elements respectively ensure local consistency of computation and post-hoc corroboration of audits. Their mathematical characterization is provided in Sec.~\ref{sec:sim_model}.

A naive solution is to validate each step (iteration) of the process using independent recomputation by validating agents. Similarly, the integrity of the computational process can be verified from an immutable record of validated intermediate states. However, practical simulations are long and involve a large number of iterations. Validation requires communication of the iterates to the endorsers, and recording the validated state on the immutable data structure. This results in significant communication and storage overhead if every state is reported and stored as is, in addition to the computational cost of validation, preventing its adoption to large-scale systems and computational methods.

It is thus important to utilize the underlying structure of the simulation to reduce these overheads. This can be done by reporting a compressed version of the states with sufficient detail such that they can be validated to within a desired tolerance. We use universal compressors to reduce these communication and storage costs. Each block of communication and storage also incurs the overhead corresponding to headers and metadata. It is thus prudent to combine multiple iterates into a single frame before compression, and collectively validate and store frames of the computational process. 

Blockchain systems establish trust in transactional systems for peer-to-peer networks of agents through distributed endorsements, consensus on transactional validity, and the storage of the collection of all transactions in the network in a shared, append-only, immutable, distributed ledger at each peer in the network. We leverage these features directly (1) to use blockchain transactions to record steps of the computation, and (2) to facilitate the immutable storage of validated audits.

Allowing validation and verification of computations not only creates an environment of trust among agents, but also enforces a higher degree of conformation and consistency in experiments. Necessitating validation and verification also implies a shared common mechanism for model and data sharing, enabling scientific reproducibility. The setup also facilitates well-defined processes for distributed and derived computing, wherein the former involves a computational framework performed piecewise at multiple nodes, and the latter concerns deriving new experiments using checkpoints drawn from the intermediate audits of prior computational experiments.

\section{Computation and Trust Model} \label{sec:sim_model}

Let us now mathematically formalize the computation model, and validation and verification requirements under consideration. We consider an iterative computational algorithm in this paper.

Consider a computational process that updates a system state, $X_t \in \reals^d$, over iterations $t \in \sth{1,2,\dots}$, depending on the current state and an external source of randomness $\theta_t \in \reals^{d'}$, according to an atomic operation $f: \reals^{d} \times \reals^{d'} \to \reals^d$ as follows
\begin{equation} \label{eqn:iter_update}
X_{t+1} = f(X_t,\theta_t).
\end{equation} 
For simplicity we assume that $\theta_t$ is shared by all agents. This can easily be generalized as elaborated later. We also assume that the function $f(\cdot)$ is $L$-Lipschitz continuous in the system state, without loss of generality, under the Euclidean norm, for all $\theta \in \reals^{d'}$ i.e.,
\begin{equation}\label{eqn:Lipschitz_cont}
\eunorm{f(X_1,\theta) - f(X_2,\theta)} \leq L \eunorm{X_1 - X_2}.
\end{equation}
That is, minor modifications to the inputs of the atomic operation result in correspondingly bounded variation in the outputs. This is expected for instance in simulations of physical or biological processes, as seen in epidemiological and meteorological simulations, as most physical systems governing behavior in nature are smooth. For instance, with respect to the OpenMalaria example, the requirement implies that minor changes in policies result in minor changes in outcomes.

We consider a multi-agent system where one agent, referred to as the \emph{computing client} (client in short), runs the computational algorithm. The other agents in the system, called \emph{peers}, are aware of the atomic operation $f(\cdot)$ and share the same external randomness and hence can recompute the iterations. Validation of intermediate states is performed by independent peers referred to as \emph{endorsers} through an iterative recomputation of the reported states from the the most recent validated state using the atomic operation $f(\cdot)$. The process of validation is referred to as an \emph{endorsement} of the state. A reported state, $\tilde{X}_t$ is \emph{valid} if it lies within a margin, $\Delta_{\text{val}}$, of the state $\hat{X}_t$ as recomputed by the endorser, i.e.,
\begin{equation} \label{eqn:validity_condn}
\eunorm{\tilde{X}_t - \hat{X}_t} = \eunorm{\tilde{X}_t - f(\tilde{X}_{t-1})} \leq \Delta_{\text{val}}.
\end{equation}
The validation criterion \eqref{eqn:validity_condn}, without loss of generality, associates equal weightage to each component of the state, and can be easily generalized to weighted norms or other notions of distance.

Verification concerns checking for integrity of the computational process which is enabled through the storage of frequent audits of validated states. Thus, if the audits record the states $\sth{\tilde{Y}_1,\tilde{Y}_2,\dots}$, then verification corresponds to ensuring that the recomputed version, $\hat{Y}_t$, of the state is within a margin, $\Delta_{\text{ver}}$, of the recorded version, i.e.,
\begin{equation} \label{eqn:ver_req}
\eunorm{\hat{Y}_t - \tilde{Y}_t} \leq \Delta_{\text{ver}}.
\end{equation}
Without loss of generality, validation requirements are stricter than for verification, i.e., $\Delta_{\text{val}} \geq \Delta_{\text{ver}}$. We now construct the system to address these two trust requirements.

\section{Multi-Agent Blockchain Framework} \label{sec:MBF}

We elaborate the system design, starting with the functional categorization of the network. We then elaborate each functional unit, including the compression at the client, the validation by endorsers, and the role of orderers in adding blocks to the ledger. For ease, let us consider a deterministic iterative algorithm for computation, $X_{t+1} = f(X_t)$.

\subsection{Peer-to-Peer Network---Functional Decomposition}

\begin{figure}[t]
	\centering
	\includegraphics[scale=0.4]{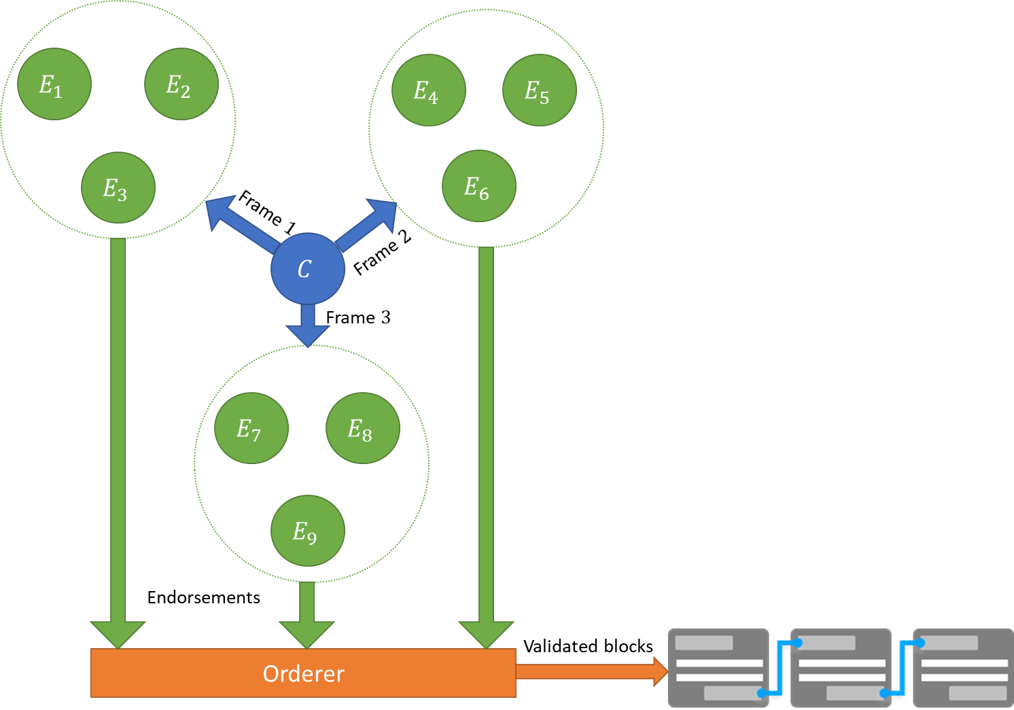}
	\caption{Functional categorization of peer-to-peer network: Clients run the iterative algorithm; multiple independent frames are validated in parallel by non-overlapping subsets of endorsers; orderers check consistency and append valid frames to the blockchain.}
	\label{fig:peer_network_config}
\end{figure}

The peer-to-peer network is functionally categorized into clients, endorsers, and orderers, who function together in computing, validating, ordering, and storing the simulated states on the blockchain ledger. Their functioning is as follows:
\begin{enumerate}
\item The client runs the computations and iteratively computes the states $X_t$, for $t \geq 1$.
\item The client groups a sequence of states into a frame, compresses, and communicates the frame to a set of endorsers.
\item The endorsers decompress frames, validate states by recomputing them iteratively, and report endorsements to orderers.
\item The orderers subsample and add the frame to the blockchain if it has been validated, and if all prior frames have been validated and added to the ledger.
\item The peers update their copy of the ledger.
\end{enumerate}
This is depicted in Fig.~\ref{fig:peer_network_config}. The classification is only based on function and each peer can perform different functions across time. Since states are grouped into independent frames, they can be validated by non-overlapping subsets of endorsers in parallel.

\subsection{Client Operations}

Clients performs the computations, compute the states, construct frames of iterates, compress, and report them to endorsers. We assume there exists an endorser assignment policy.

Owing to the Lipschitz continuity,
\[
\eunorm{X_{t+1} - X_t} \leq L \eunorm{X_t - X_{t-1}}.
\]
Thus state updates (differences) across iterates are bounded to within a factor of the deviation in the previous iteration. This property can be leveraged to compress state updates using delta encoding \cite{GrangerJ1980}, where states are represented in the form of differences from the previous state. Then, it suffices to store the state at certain checkpoints of the computational process, with the iterates between checkpoints represented by the updates. 

We describe the construction inductively, starting with the initial state $X_0$, assumed to be the first checkpoint. Let us assume that the state reported at time $t$ is $\tilde{X}_t$ and the true state is $X_t$. Then, if $X_{t+1} = f(X_t)$, define the update as 
\[
\Delta X_{t+1} = X_{t+1} - \tilde{X}_t.
\] 
The cost of communication (for validation) and storage (for verification) of these updates is reduced by performing lossy compression (vector quantization \cite{GershoG2012}). Let the quantizer be represented by $Q(\cdot)$ and let the maximum quantization error magnitude be $\epsilon$, i.e., if the client reports $\tilde{\Delta} X_t = Q(\Delta X_t)$, then,
\begin{equation} \label{eqn:quantization_error}
\eunorm{\tilde{\Delta} X_t - \Delta X_t} \leq \epsilon.
\end{equation}

Additionally, the checkpoints can also be compressed using a Lempel-Ziv-like dictionary-based lossy compressor. Here, a dictionary of unique checkpoints are maintained. For each new checkpoint, we first check if the state is within a margin $\epsilon$ from an entry in the dictionary, and the index of this entry is reported. If not, the state is added to the dictionary and its index is reported. Other universal vector quantizers can also be utilized for compressing checkpoints, and we denote this quantizer by $\tilde{Q}(\cdot)$.

\begin{figure}[t]
	\centering
	\includegraphics[scale=0.35]{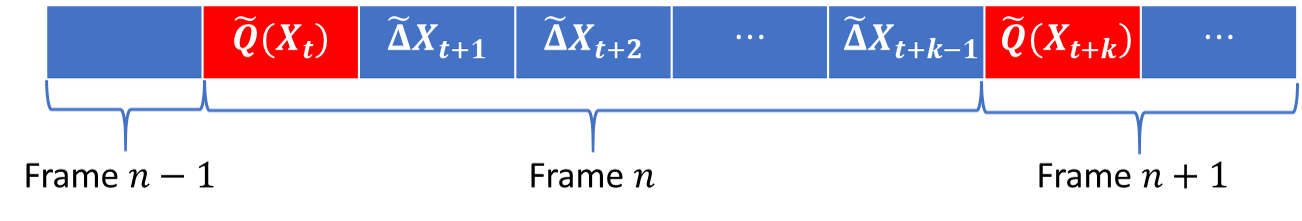}
	\caption{Structure of frames: Each frame includes a header followed by compressed updates of successive iterates.}
	\label{fig:frame_const}
\end{figure}

Let $\Delta_{\text{quant}}$ be the maximum magnitude of a state update within a frame, i.e., if $\eunorm{\Delta X_t } > \Delta_{\text{quant}}$, the client creates a checkpoint at $t+1$ and reports $\tilde{X}_{t+1} = \tilde{Q}(X_{t+1})$. Then $X_{t+1}$ is reported as
\begin{equation}
\tilde{X}_{t+1} = \begin{cases}
\tilde{Q}(X_{t+1}) &, \text{ if } t+1 \text{ is a checkpoint} \\
\tilde{X}_t + \tilde{\Delta} X_{t+1} &, \text{ o/w }
\end{cases}.
\end{equation}
The resulting sequence of frames is as shown in Fig.~\ref{fig:frame_const}. 

\begin{figure}[t]
	\centering
	\includegraphics[scale=0.4]{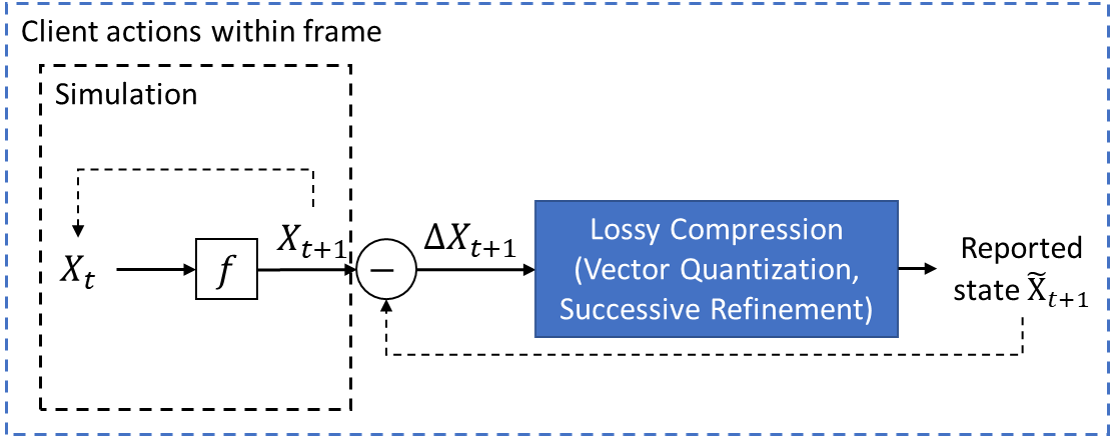}
	\caption{Operations performed by the client within a frame. The client computes the states according to the iterative algorithm, performs delta encoding, and communicates the compressed state updates to the endorsers in frames.}
	\label{fig:client_ops}
\end{figure}

Separate from creating new checkpoints adaptively, the system can also restrict the maximum size of a frame by a constant $\bar{M}$ to limit the computational overhead of its validation. Fig.~\ref{fig:client_ops} summarizes the tasks performed by the computing client.

The choice of design parameters, $\epsilon, \Delta_{\text{quant}}$, are to be made such that the reports are accurate enough for validation. The optimal design choice is shown in the following results.

\begin{theorem} \label{thm:quant_criterion}
If $f(\cdot)$ is $L$-Lipschitz continuous, and $\epsilon \leq \tfrac{\Delta_{\text{val}}}{L+1}$, then, a state $\tilde{X}_t$ is invalidated by an honest endorser only if there is a computational error of magnitude at least $\epsilon$, i.e., $\eunorm{\tilde{X}_t - X_t} \geq \epsilon$.
\end{theorem}
\begin{IEEEproof}
Let us assume that $\tilde{X}_t$ is a valid state. Then,
\begin{align}
\eunorm{\hat{X}_{t+1} - \tilde{X}_{t+1}} &\leq \eunorm{\hat{X}_{t+1} - X_{t+1}} + \eunorm{X_{t+1} - \tilde{X}_{t+1}} \notag \\
&\leq L\eunorm{\tilde{X}_{t} - X_{t}} + \eunorm{\tilde{\Delta} X_t - \Delta X_t} \label{eqn:Lipschitz_pf} \\
&\leq L \eunorm{\tilde{\Delta} X_{t-1} - \Delta X_{t-1}} + \epsilon \label{eqn:state_update_def} \\
&\leq (L+1) \epsilon \leq \Delta_{\text{val}}, \label{eqn:eps_condn}
\end{align}
where \eqref{eqn:Lipschitz_pf} follows from the Lipschitz continuity of the atomic operation, \eqref{eqn:state_update_def} is from the definition of the compressed state updates, and \eqref{eqn:eps_condn} follows from the quantization error bound.
\end{IEEEproof}
\begin{corollary} \label{cor:suff_condn}
If $\eunorm{\tilde{X}_t-X_t} \geq \Delta_{\text{val}} + L\epsilon$, then $\tilde{X}_t$ is invalidated.
\end{corollary}
The necessary and sufficient conditions for invalidation in Thm.~\ref{thm:quant_criterion} and Cor.~\ref{cor:suff_condn} highlight the fact that computational errors of magnitude less than $\epsilon$ are missed, and any error of magnitude at least $\Delta_{\text{val}} + L\epsilon$ is certainly detected. When the approximation error is made arbitrarily small, all errors beyond the tolerance are detected. A variety of vector quantizers, satisfying Thm.~\ref{thm:quant_criterion} can be used for lossy delta encoding---one simple choice is lattice vector quantizers \cite{ServettoVS1999}.
\begin{theorem}
Let $\calB(\Delta_{\text{quant}}) = \sth{x\in\reals^d : \eunorm{x} \leq \Delta_{\text{quant}}}$ and let $\Delta X_t \sim \text{Unif}(\calB(\Delta_{\text{quant}}))$, then the communication and storage cost per state update is $O\pth{d \log\pth{\tfrac{\Delta_{\text{quant}}}{\epsilon}}}$ bits.
\end{theorem}
This follows directly from the covering number of $\calB(\Delta_{\text{quant}})$ using $\calB(\epsilon)$ balls; a similar cost is incurred for other standard lattices.

\begin{theorem}
For any frame $n$, with checkpoint at $T_n$, maximum number of states in the frame, $M_n$, is bounded as
\begin{equation} \label{eqn:frame_size_LB}
M_n \geq \min\sth{\frac{\log \pth{\Delta_{\text{quant}} - \epsilon} - \log \delta_n}{\log L}, \bar{M}},
\end{equation}
where $\delta_n = \eunorm{X_{T_n + 1} - X_{T_n}}$, is the first update in the frame.
\end{theorem}
\begin{IEEEproof}
Without loss of generality, let us consider the first frame, i.e., $n=1, T_n = 0$. Then, $M_n = t$ implies that
\begin{align}
\Delta_{\text{quant}} &\leq \eunorm{\Delta X_t} = \eunorm{X_{t+1} - \tilde{X}_t} \notag \\
&\leq \eunorm{X_{t+1} - X_t} + \eunorm{X_t - \tilde{X}_t} \label{eqn:FS_LB_tri_ineq} \\
&\leq L^t \eunorm{X_1 - X_0} + \epsilon = L^t \delta_1 + \epsilon,\label{eqn:FS_LB_Lips_quant} 
\end{align}
where \eqref{eqn:FS_LB_tri_ineq} follows from the triangle inequality, and \eqref{eqn:FS_LB_Lips_quant} follows from the Lipschitz continuity and the quantization error. Thus the result follows, for any frame, by direct extension.
\end{IEEEproof}
This provides a simple sufficient condition on the size of a frame, in terms of the magnitude of the first iteration in the frame. Naturally a small first iterate implies the possibility of accommodating more iterates in the frame. This lower bound can be used in identifying the typical frame size and the corresponding costs of communication and computation involved, prior to the design of the scheme. We describe the generalization of the compressor to the parameter unaware setting in Sec.~\ref{sec:extensions}.

\subsection{Endorser and Orderer Operations}

\begin{figure}[t]
	\centering
	\includegraphics[scale=0.4]{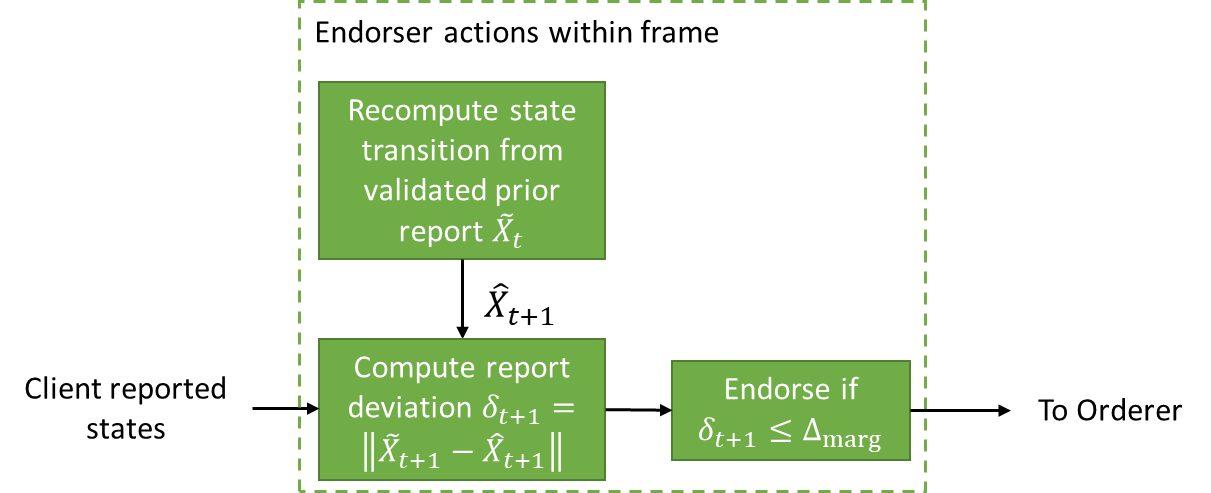}
	\caption{Operations performed by the endorser for a single frame: The endorser sequentially validates the states reported by the client by decoding the updates and recomputing the state from the atomic operation.}
	\label{fig:endorser_ops}
\end{figure}

We now define the role of an endorser in validating a frame. A summary of the operations is depicted in Fig.~\ref{fig:endorser_ops}. For preliminary analysis, we assume that endorsers are honest and are homogeneous in terms of communication latency and computational capacity. A more refined allocation policy can be designed to account for the case of variabilities in communication and computational costs. However we do not consider this in this paper.

Each endorser involved in validating a frame, sequentially checks the state updates by recomputing from the last valid state, i.e., to validate the report $\tilde{X}_{t+1}$, the endorser computes $\hat{X}_{t+1} = f(\tilde{X}_t)$ and checks for the validity criterion \eqref{eqn:validity_condn}. The frame is reported as valid if all updates are valid in the frame. The endorsements are then reported to the orderer.

Individual update validations can also be performed in parallel and finally verified for sequential consistency. Such parallelism can be performed either at the individual endorser-level, or in the form of the distribution of the sub-frames across endorsers. This results in a reduction of the time required for validating a frame. For the sake of simplicity, we skip these extensions in this paper.

\begin{figure}[t]
	\centering
	\includegraphics[scale=0.4]{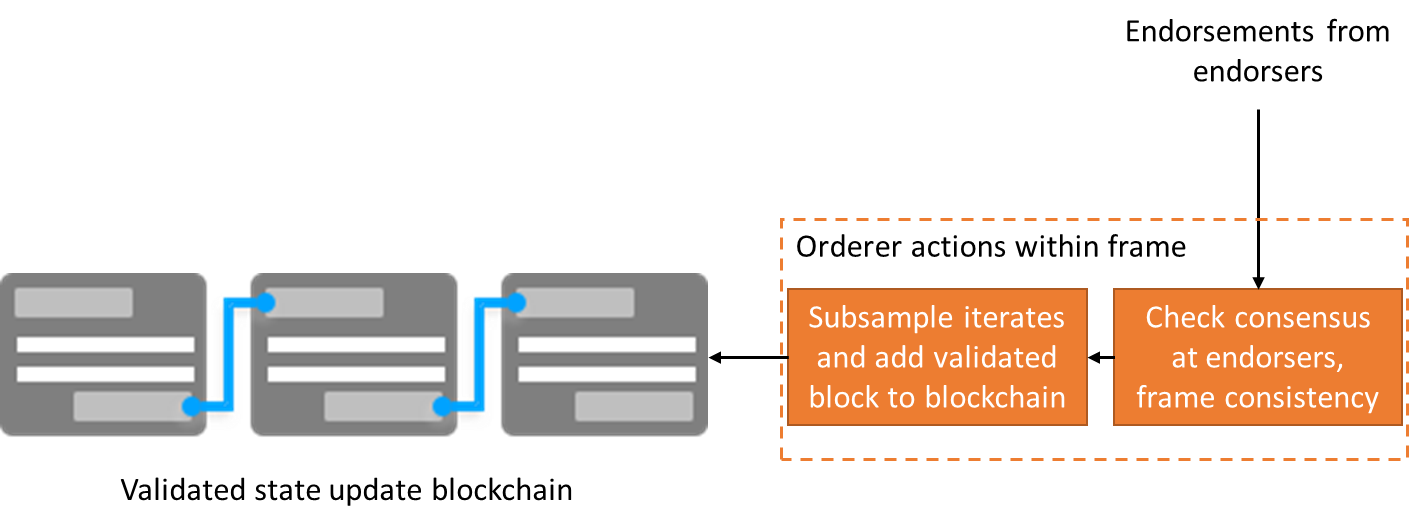}
	\caption{Operations performed by the orderer for frames. The orderer sequentially adds valid frames to the blockchain after checking for consistency.}
	\label{fig:orderer_ops}
\end{figure}

Upon receiving the endorsements for frames, the orderer checks for consistency of the checkpoints and adds a valid frame to the blockchain ledger if all prior frames have already been added, and broadcasts the block to other peers. This is depicted in Fig.~\ref{fig:orderer_ops}.

Since the state updates are stored on the immutable data structure of the blockchain, they provide an avenue for verification of the computations at a later stage. As described in \eqref{eqn:ver_req}, the verification requirements are not as strict as the validation requirements. Thus it suffices to  subsample the updates in a frame and store only a subset, i.e., one state is stored for every $K = \tfrac{\Delta_{\text{ver}}}{\Delta_{\text{val}}}$ iterates. Then, the effective state update is the sum of the individual updates of the $K$ intermediate iterates.

A block stored on the blockchain is now characterized by the audits that are either the checkpoints or the cumulative updates corresponding to $K$ successive iterates. The audits $\tilde{Y}_{\tau}$ are then defined by
\begin{equation} \label{eqn:subsample_storage}
\tilde{Y}_{\tau + 1} = \begin{cases}
\tilde{Y}_{\tau} + \sum \tilde{\Delta} X_{t+1} & , \text{ if no checkpoint in  next $K$ iterates} \\
\tilde{X}_{t'} & , \text{ otherwise }
\end{cases},
\end{equation}
where the sum is over the intermediate iterates, and $t'$ is the next checkpoint. Then, the audits are grouped into blocks as described by Fig.~\ref{fig:frame_const} and added to the blockchain ledger by the orderer.

\begin{theorem}
For subsampled storage at a frequency $1/K$ according to \eqref{eqn:subsample_storage}, a Lipschitz constant $L$ of $f(\cdot)$, and quantization error $\epsilon$,
\begin{equation} \label{eqn:verif_dev}
\eunorm{\hat{Y}_{\tau+1} - \tilde{Y}_{\tau+1}} \leq \pth{L^K + 1} K \epsilon,
\end{equation}
where $\hat{Y}_{\tau + 1} = f^K(\tilde{Y}_{\tau})$.
\end{theorem}
\begin{IEEEproof}
First, $f^K(\cdot)$ is $L^K$-Lipschitz continuous. Then,
\begin{align}
\eunorm{\hat{Y}_{\tau+1} - \tilde{Y}_{\tau+1}} &\leq \eunorm{f^K(\tilde{Y}_{\tau}) - f^K(Y_{\tau})} + \eunorm{Y_{\tau+1} - \tilde{Y}_{\tau+1}} \label{eqn:ver_dev_tri_ineq} \\
&\leq L^K  \eunorm{\tilde{Y}_{\tau} - Y_{\tau}} + \eunorm{\sum_{t \in \calT_{\tau}} \pth{\tilde{\Delta} X_t - \Delta X_t}} \label{eqn:ver_dev_Lips_def} \\
&\leq (L^K + 1) K\epsilon, \notag
\end{align}
where \eqref{eqn:ver_dev_tri_ineq} follows from the triangle inequality, and \eqref{eqn:ver_dev_Lips_def} follows from the Lipschitz inequality and \eqref{eqn:subsample_storage}.
\end{IEEEproof}
Thus, a viable subsampling frequency can be determined by finding a $K$ such that $(L^K+1) K \leq \tfrac{\Delta_{\text{ver}}}{\epsilon}$. This reduces the storage cost on the blockchain at the expense of accuracy of recorded audits. If the agents are interested in increasing accuracy of the records over time, then the quantizers can be dynamically adjusted accordingly.

\subsection{Example Application} \label{sec:example}

Let us now elaborate the design from the context of a synthetic example that is used later for experimental study in Sec.~\ref{sec:experiments}. Consider an agent, the client, in a network who wishes to address a simple classification problem using training data that he has access to. The client aims to train a neural network using backpropagtion based on mini batch stochastic gradient descent (SGD) to solve this classification problem and subsequently share the trained network with other agents who are also interested in the solution.

However, the client is limited by the amount of computational resources available for training, and also does not wish to share the private data used for training. Since it has limited resources toward gradient computation, it uses small batch sizes to get faster estimates. However, the peers do not trust the computations performed by the client, not just because of its proclivity toward errors arising from computational limitations, but also possible malicious intent. The peers themselves have access to private datasets, drawn from the same source, but much smaller in size such that they can not train a network on their own for the task.

In such a context, we can establish distributed trust in the agents using MBF as follows:
\begin{enumerate}
\item The client sets up the training with parametric inputs (network architecture, learning rates, batch sizes etc.) and shares them over the blockchain with other peers.
\item The client runs the training algorithm, compresses state updates (network weights) using lossy compression and reports compressed frames to endorsers for validation. 
\item The endorsers rerun steps of mini batch SGD from validated prior reports. They compute deviation (Fig.~\ref{fig:endorser_ops}) in network weights, endorse according to \eqref{eqn:validity_condn}, and communicate validated frames and endorsements to the orderer.
\item MBF orderer checks for consensus among endorsers and subsamples the frame to construct blocks. They then add blocks sequentially to the blockchain ledger.
\item Client reports the network to peers at the end of training.
\end{enumerate}
Since the experiment is run on the MBF platform, peers are assured of validity of steps of the training, and also have access to the blockchain to verify the computations. Since the private training data is not shared across peers, the endorsement process for revalidation needs to be appropriately adjusted. This is described in Sec.~\ref{sec:extensions}, and a detailed experimental study of this problem, adapted to the MNIST dataset, is done in Sec.~\ref{sec:experiments}.

Thus the MBF platform addresses the trust issues described in Sec.~\ref{sec:motivation} and allows for efficient collaboration and trusted data, model, and result sharing among agencies involved in malaria research and policy design.

\section{Design Advantages and Costs} \label{sec:design_costs}

We now perform a cost-benefit analysis of the design. To the best of our knowledge this is the first system designed to address trust in such systems and so we benchmark the costs against simpler implementations to emphasize the importance of the different components of the system.

Let us first identify the advantages of the platform.
\begin{itemize}
\item {\bf Accountability:} The MBF platform guarantees provenance through the immutable record of computations. Thus, we can not only detect the source of potential conflicts, but also to trace ownership of computations.
\item {\bf Transparency:} The platform establishes trust among agents through a transparent record of the validated computational trajectories of computation.
\item {\bf Adaptivity:} The frame design, endorsement, and validation methods adapt according to the state evolution. Further, the validity margins can be altered across time by dynamically varying the quantizers. In convergent simulations/algorithms, the system can thus use monotonically decreasing margins to obtain stricter guarantees at convergence.
\item {\bf Generality:} The platform uses fairly general building blocks, and can be easily implemented using existing methods.
\item {\bf Computation universality:} The design is agnostic to computational process specifics and can be implemented as long as it is composed of reproducible atomic computations.
\item {\bf Scientific reproducibility:} By storing intermediate states this platform guarantees reliable data and model sharing, and collaborative research. It thus facilitates scientific reproducibility in large-scale computational experiments.
\end{itemize}

To compare the costs of the system, let us consider three different modes for such blockchain-based distributed trust:
\begin{enumerate}
\item {\bf Transaction Mode:} Here we treat each iteration as a transaction and validate and store each state transition as a block on the blockchain ledger independently.
\item {\bf Streaming Mode:} Here each state is independently compressed according to a universal compressor, validated, and stored on the blockchain.
\item {\bf Batch Mode:} This corresponds to the MBF design described in this paper.
\end{enumerate}

Let us assume that the average number of endorsers per frame be $\bar{E}$, the average size of the frame be $\bar{M}$, and the subsampling frequency be $\nu = 1/K$ in the batch mode. We benchmark costs relative to this average set of $\bar{M}$ iterations, and the same computational redundancy.

First, let us consider the computational overhead involved. Each mode performs $(1+E)$-times as many computations as the untrusted simulation. The streaming and batch modes additionally incur the cost of compression and decompression of states. The batch mode also includes the cost of subsampling the frames. Thus we can see that the transaction mode incurs the least computational overhead, while the batch mode incurs the most. Informally, the batch mode incurs a cost of
\begin{equation}
C^{(\text{comp.})}_{\text{batch}} = (1+E) C^{(\text{comp.})}_\text{sim} + C^{(\text{comp.})}_\text{compression} + C^{(\text{comp.})}_\text{sampling},
\end{equation}
where $C^{(\text{comp.})}_{\text{sim}}, C^{(\text{comp.})}_{\text{compression}}, C^{(\text{comp.})}_{\text{sampling}}$ are the computational costs of the simulation, compression and decompression, and subsampling respectively. The transaction and streaming modes incur just the first and the first two costs respectively.

The communication overheads include the state reports and metadata used for validation and coordination respectively. In the transaction mode, as states are uncompressed, the communication cost is significant and is not scalable. On the other hand, the streaming and batch modes reduce these costs through lossy compression. Assuming a bounded set of states, $\calX$, such that $\max_{X\in\calX} \eunorm{X} = B \gg \Delta_{\text{quant}}$, the worst-case sufficient communication cost in transaction mode using vector quantization for $\bar{M}$ iterations is
\begin{equation}
C^{(\text{comm.})}_{\text{transaction}} = O\pth{\bar{M} d \log_2\pth{\frac{B}{\epsilon}} + \bar{M} C^{(\text{comm.})}_{\text{meta}}},
\end{equation}
where $C^{(\text{comm.})}_{\text{meta}}$ is the average communication cost of metadata, per instance of communication. On the other hand, the batch mode reduces both compression cost, and the metadata, as
\begin{equation}
C^{(\text{comm.})}_{\text{batch}} = O\pth{\bar{M} d \log_2\pth{\frac{\Delta_{\text{quant}}}{\epsilon}} + d \log_2\pth{\frac{B}{\epsilon}} + C^{(\text{comm.})}_{\text{meta}}}.
\end{equation}
The costs expressed are sufficient costs in the order sense and more precise estimates can be computed given the compression scheme and statistics of state evolution.

Similarly, with regard to storage, the transaction mode incurs significant costs on account of not compressing the audits. The batch mode not only incurs lesser metadata for storage but also fewer state updates on account of subsampling when compared to the streaming mode. To be precise, $C^{(\text{storage})}_{\text{transaction}} = C^{(\text{comm.})}_{\text{transaction}}$ in the order sense, whereas
\begin{equation}
C^{(\text{storage})}_{\text{batch}} = O\pth{\nu \bar{M} d \log_2\pth{\frac{\Delta_{\text{quant}}}{\epsilon}} + d \log_2\pth{\frac{B}{\epsilon}} + C^{(\text{storage})}_{\text{meta}}}.
\end{equation}

Thus the batch mode reduces communication and storage overheads at the expense of added computational cost. Through a careful analysis of the tradeoffs, we can adopt optimal compression schemes and subsampling mechanisms.

\section{Extensions of Design} \label{sec:extensions}

We now describe a couple of avenues for generalization.

\subsection{Parameter Agnostic Design}

In Sec.\ref{sec:MBF} we used a vector quantizers based on the Lipschitz constant $L$. In practice, such parameters of the computational algorithm are not known \emph{apriori}. Underestimating $L$ can result in using a larger quantization error, that could cause errors in validation even when the client computes correctly. In such cases, it is essential to be able to identify the cause for the error.

One option is to estimate $L$ from computed samples. This translates to estimating the maximum gradient magnitude for the atomic operation which might be expensive in sample and computational complexity, depending on the application. Thus, we propose an alternative compression scheme.

We draw insight from video compression strategies, and propose the use of successive refinement coding \cite{EquitzC1991} of the state updates. That is, a compression bit stream is generated for each state update such that the accuracy can be improved by sending additional bits from the bit stream. Successive refinement allows the endorsers to provide updates on the report such that the state accuracy can be iteratively improved.

Thus, if an invalidation notice is received from endorsers, the client has two options---checking the computations, and/or refining the reported state through successive refinement. Depending on the computation-communication cost tradeoff, the client appropriately chooses the more economical alternative. Through successive refinement, the client provides more accurate descriptions of the state vector, and thus reduces the possibility of validation errors caused by report inaccuracy.

One possible efficient compression technique uses lattice vector quantizers \cite{MukherjeeM2002,LiuP2007} to define successive refinement codes. This also reduces the size of the codebook, if the refinement lattices are assumed to be of the same geometry, because the client only needs to communicate the scaling corresponding to the refinement. This allows for improved adaptability in the refinement updates. More efficient quantizers can also be defined if additional information regarding the application and state updates are available.

\subsection{Computations with External Randomness}

As described in Sec.~\ref{sec:sim_model}, such computational algorithms in practice typically evolve iteratively as a function of the current state $X_t$, and an external randomness $\theta_t$. When this randomness is not shared across agents, and is inaccessible to the client, reproduction of the reported results by an endorser becomes infeasible and so is validating local computations. This could also emerge in cases where the client is unwilling to share private data associated used by the algorithm with other agents \cite{VermaCC2018}.

For instance, in simulations of disease spread using black box models, each run of the simulation adopts a different outcome, depending on the underlying random elements introduced by the model to mimic societal and pathological disease spread factors. Quite often, the client does not have access to all the random elements introduced by the model in creating that particular outcome.

Whereas the exact random instance might not be available, the source of such randomness is often common, i.e., $\theta_t \sim P_{\theta}$, and $P_{\theta}$ is known. In this context, we redefine validation as guaranteeing \eqref{eqn:validity_condn} with probability at least $1-\rho$, i.e.,
\begin{equation} \label{eqn:validity_condn_rand}
\prob{\eunorm{\tilde{X}_t - \hat{X}_t} \geq \Delta_{\text{marg}}} \leq \rho. 
\end{equation}
This requirement removes outliers in the computation process and only allows trajectories close to the expected behavior.

Then, we can exploit the law of large numbers to validate reports by their deviation from the average behavior observed across multiple independent endorsers, 
\[
\hat{X}_{t+1} = \frac{1}{m} \sum_{i=1}^m f(\tilde{X}_t, \theta_i),
\]
where $\theta_i \stackrel{\text{i.i.d.}}{\sim} P_{\theta}$. By choosing a sufficiently large number of endorsers, depending on $\rho$, we can assure \eqref{eqn:validity_condn_rand}. The role of the endorsers is appropriately modified and the system calls for higher coordination amongst the endorsers. 

Using multi-variate concentration inequalities, we can also quantify the sufficient number of endorsers for validation.
\begin{theorem} \label{thm:prob_dev_cheby_bd}
Let $\epsilon < \tfrac{\Delta_{\text{marg}}}{L+1}$. For a state at time $t$, if the average of $m$ endorsers is used for validation,
\begin{equation} \label{eqn:dev_prob_rand}
\prob{\eunorm{\tilde{X}_t - \hat{X}_t} \geq \Delta_{\text{marg}}} \leq \frac{2d\tilde{\lambda}^2}{(\Delta_{\text{marg}} - (L+1)\epsilon)^2} \pth{1+\frac{1}{m\tilde{\lambda}}}^2,
\end{equation}
where $\tilde{\lambda}$ is the maximum eigenvalue of covariance matrix of of the quantized state vector.
\end{theorem}
\begin{IEEEproof}
For a $d$-dimensional random variable $X$ with $\expect{X} = \mu, \expect{(X-\mu)(X-\mu)^T} = V$, according to the multidimensional Chebyshev inequality \cite{MarshallO1960},
\[
\prob{\eunorm{X-\mu}_{V^{-1}} >t} \leq \frac{d}{t^2}.
\]
Then, using the fact that $\lambda_{min} \eunorm{x} \leq \eunorm{x}_A \leq \lambda_{max} \eunorm{x}$, for any vector $x$ and matrix $A$ with minimum and maximum eigenvalues $\lambda_{min}, \lambda_{max}$, we have
\begin{equation} \label{eqn:multi_cheby}
\prob{\eunorm{X-\mu} >t} \leq \frac{\lambda^2 d}{t^2},
\end{equation}
where $\lambda$ is the maximum eigenvalue of $V$.

Further, from the bound on the quantization error, we can observe that for $\tilde{X}_{t+1} = X_{t+1} + Z_t$, where $Z_t \in \calB(\epsilon)$. Then,
\[
\eunorm{\expect{X_{t+1}} - \expect{\tilde{X}_{t+1}}} \leq \epsilon.
\]
Then,
\begin{align}
& \eunorm{\expect{\tilde{X}_{t+1}} - \expect{\hat{X}_{t+1}}} = \eunorm{\expect{\tilde{X}_{t+1}} - \expect{f(\tilde{X}_t,\theta)}} \notag\\
&\quad \leq \eunorm{\expect{X_{t+1}} - \expect{\tilde{X}_{t+1}}} + \eunorm{\expect{f(X_t,\theta-f(\tilde{X}_t,\theta)}} \label{eqn:expect_tri_ineq} \\
&\quad \leq \epsilon + L \eunorm{\expect{X_{t}} - \expect{\tilde{X}_{t}}} \label{eqn:quant_bd} \\
&\quad \leq (L+1)\epsilon, \label{eqn:expect_bd}
\end{align}
where \eqref{eqn:expect_tri_ineq} follows from the triangle inequality, and \eqref{eqn:quant_bd} follows from the bound on the quantization error, the fact that $\expect{\eunorm{X}} \geq \eunorm{\expect{X}}$, and the Lipschitz continuity.

Finally, for any $\alpha \in (0,1)$,
\begin{align}
&\prob{\eunorm{\tilde{X}_{t+1} - \hat{X}_{t+1}} \geq \Delta_{\text{marg}}} \notag \\
&\quad \leq \prob{\eunorm{\tilde{X}_{t+1} - \expect{\tilde{X}_{t+1}}} \geq \alpha (\Delta_{\text{marg}} - (L+1)\epsilon)} \notag \\
&\qquad + \prob{\eunorm{\hat{X}_{t+1} - \expect{\hat{X}_{t+1}}} \geq (1-\alpha) (\Delta_{\text{marg}} - (L+1)\epsilon)} \label{eqn:union_bd_use}\\
&\quad \leq \frac{d}{(\Delta_{\text{marg}} - (L+1)\epsilon)^2} \pth{\frac{\tilde{\lambda}^2}{\alpha^2} + \frac{1}{m^2 (1-\alpha)^2}} \label{eqn:Chebyshev_ineq_use}\\
&\quad \leq \frac{2d \tilde{\lambda}^2}{(\Delta_{\text{marg}} - (L+1)\epsilon)^2} \pth{1+\frac{1}{m\tilde{\lambda}}}^2, \label{eqn:max_over_alpha}
\end{align}
where \eqref{eqn:union_bd_use} follows from the triangle inequality and the union bound, and \eqref{eqn:expect_bd}, and \eqref{eqn:Chebyshev_ineq_use} from \eqref{eqn:multi_cheby}. Finally, \eqref{eqn:max_over_alpha} is obtained by maximizing the bound over $\alpha \in (0,1)$.
\end{IEEEproof}

\begin{corollary}
To guarantee validation with probability at least $1-\rho$, for a margin of deviation of $\Delta_{\text{marg}}$, where $\rho \leq \tfrac{2d}{(\Delta_{\text{marg}} - (L+1)\epsilon)^2}$, it suffices to use 
\begin{equation} \label{eqn:no_endorsers_rand}
m \geq \qth{ \pth{ \sqrt{\tfrac{\rho}{2d}} (\Delta_{\text{marg}} - (L+1)\epsilon) - \tilde{\lambda}}}^{-1}
\end{equation}
endorsers.
\end{corollary}
This sufficient condition follows directly from Thm.~\ref{thm:prob_dev_cheby_bd}.

\section{Experiments} \label{sec:experiments}

In this section, we run some simple synthetic experiments using the MNIST database \cite{LecunBBH1998}, for the scenario described in Sec.~{\ref{sec:example}}, to understand the distributed trust environment design, the costs involved, and the benefits of the enforcement. These synthetic experiments were selected to evaluate the efficacy of our approach with a domain that is familiar, and the process of training neural networks that is common in the research community.

Let us consider a simple $3$-layer neural network, trained on the MNIST database, with $25$ neurons in the hidden layer. Consider a client training the neural network using mini-batch stochastic gradient descent (SGD), with limited resources such that, it is constrained in computing gradients and so uses a small batchsize of $10$ samples per iteration and $1000$ iterations. The average precision of such a neural network trained with gradient descent is $97.4\%$. We now wish to establish trust in the training process owing to the limited resources of the client. Whereas this configuration is far from the state of the art on the database, it does help understand the trust environment better owing to its suboptimality.

Since the training process uses stochastic gradients, exact recomputation of the iterates is infeasible. Hence, we compare deviations from the average across $m = 5$ endorsers per state for validation. We evaluate the computation and communication cost of validation as a function of the tolerance chosen for validation. Since the neural network converges to a local minimum according to SGD, we use a tolerance for iteration $t$ as $\Delta_{\text{val}}(t) = \tfrac{\Delta_{\text{max}}}{\log (t+1)}$. That is, the validation requirements are made stronger with the iterations.

We consider three main cases of the simulation:
\begin{enumerate}
\item {\bf Base case:} Compression error is less than validation tolerance, i.e., $\epsilon \leq \Delta_{\text{max}}$, and maximum frame size is $10\%$ of the total number of iterations.
\item {\bf Coarse compression:} Large compression error, i.e., $\epsilon \geq \Delta_{\text{max}}$ for at least some instances, and same base $\bar{M}$.
\item {\bf Large frames:} Same base compression error, and maximum frame size is $20\%$ of total number of iterations.
\end{enumerate}

\begin{figure}[t]
	\centering
	\includegraphics[scale=0.7]{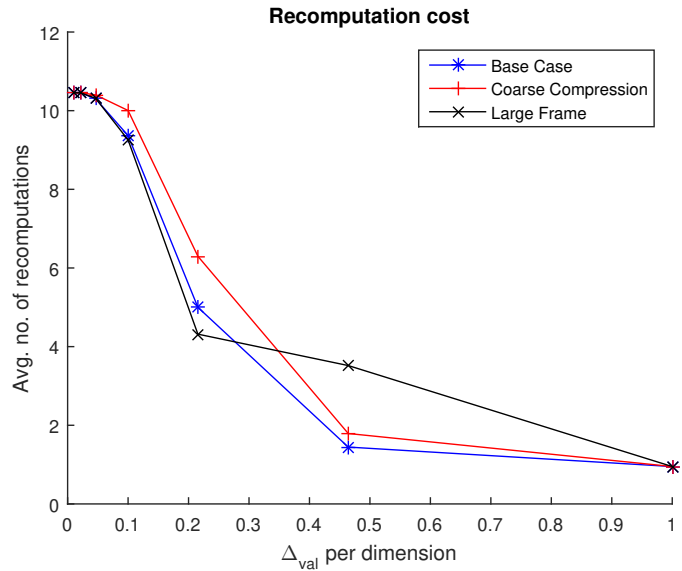}
	\caption{Average recomputation cost tradeoff: Plot depicts the average number of recomputations of gradients per iteration for varying validation requirements.}
	\label{fig:recomp_cost}
\end{figure}

\begin{figure}[t]
	\centering
	\includegraphics[scale=0.7]{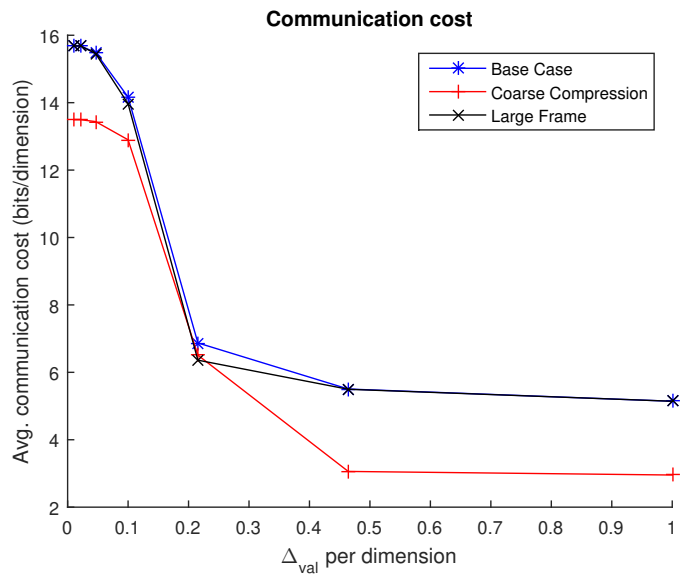}
	\caption{Average communication cost tradeoff: Plot depicts the average number of bits per dimension communicated by clients to endorsers for varying validation tolerance.}
	\label{fig:comm_cost}
\end{figure}

In the base case, invalidation from approximation errors are more frequent in later iterations when the tolerance is also lower. However, with increasing iterations, the network weights are also closer to the minima. Thus approximation errors can be eliminated by successive refinement, as gradients estimates by the client also get more accurate. The presence of outliers and smaller batch sizes impact the initial iterations much more, which are reported with comparatively better accuracy, as required by the weaker validation criterion, therein only invalidating computational errors.

In comparison, in the case of coarse compression, approximation errors of the gradients are much more likely, therein resulting in more instances of invalidation. This translates to a higher number of gradient recomputations at the expense of reduced communication overhead on the compressed state updates. On the other hand, in the case of the extremely large frames, the endorsers validate longer sequences of states at once. Thus, each invalidation results in a recomputation of all succeeding states, therein increasing the number of recomputations from the base case. This case however reduces the number of frames and checkpoints, therein reducing the average communication cost in comparison to the base case.

In Fig.~\ref{fig:recomp_cost}, the average number of gradient recomputations per iteration is shown for these three cases. As expected, this decays sharply as we increase the tolerance. Note that at either extreme, the three cases converge in the number of recomputations. This is owing to the fact that at one end all gradients are accepted whereas at the stricter end, most gradients are rejected with high probability, irrespective of the compression parameters. In the moderate tolerance range, we observe the tradeoffs as explained above. The corresponding communication cost tradeoff is shown in Fig.~\ref{fig:comm_cost}.

\begin{figure}[t]
	\centering
	\includegraphics[scale=0.7]{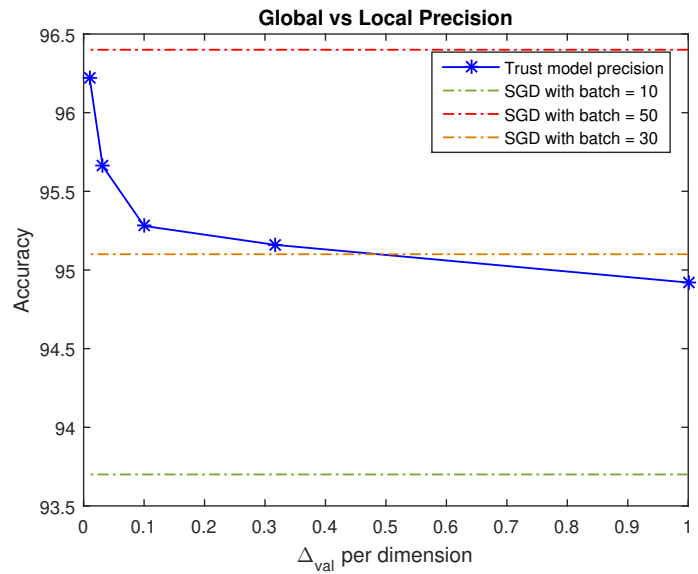}
	\caption{Precision of NN vs Trust: Plot depicts precision of the trained neural network satisfying the local validation criterion. Eliminating spurious gradients through validation enhances training process.}
	\label{fig:precision}
\end{figure}

Fig.~\ref{fig:precision} shows the precision of the neural network trained under the validation requirement as compared against the networks trained with standard mini batch SGD of batch sizes $10,30,$ and $50$. We note that the network trained with trust outperforms the case of vanilla SGD with the same batch size as it eliminates spurious gradients at validation. Increasing trust requirements (decreasing tolerance) results in improved precision of the model. In particular, it is worth noting that the strictest validation criterion results in performance that is almost as good as training with a batch size of $50$. This is understandable as the endorsers validate only those gradients that are close to that of the case with mini batch of size $50$. In fact, even when the trust requirements are fairly relaxed, just eliminating outliers in the gradients enhances the training significantly.

Thus, these simple experiments not only highlights the role of trust in guaranteeing local and global consistency in the computational process, but also the cost tradeoffs involved in establishing them. For this application for instance, appropriate parameters can be chosen by studying the precision-cost tradeoffs. Other applications might invoke similar tradeoffs, implicit or explicit, in the trust guarantees and the resulting cost of implementation.

\section{Conclusion}

In this paper we considered a multi-agent computational platform and the problem of establishing trust in the computations performed by individual agents in such a system. Using a novel combination of blockchains and distributed consensus through recomputation, we assured validity of local computations and simple verification of computational trajectories. Using efficient, universal compression techniques, we also identified methods to reduce the communication and storage overheads concerned with establishing trust in such systems, therein addressing the scalability challenge posed by blockchain systems.

Creation of such trusted platforms for distributed computation among untrusting agents allows for improved collaboration, and efficient data, model, and result sharing that is critical to establish efficient policy design mechanisms. Additionally they also result in creating unified platforms for sharing results, and in ensuring scientific reproducibility.

\section*{Acknowledgment}
This work was conducted under the auspices of the IBM Science for Social Good initiative. The authors thank Aleksandra Mojsilovi\'c and Komminist Weldemariam for discussions and support.

%%%%%%%%%%%%%%%%%%%%%%%%%%%%%%%%%%%%%%%%%%%%%%%%%%%%%%%%%%%%%%%%%%%%%%%%%%%%%%%%
\bibliographystyle{IEEEtran}
\bibliography{main}

\begin{thebibliography}{10}
\providecommand{\url}[1]{#1}
\csname url@samestyle\endcsname
\providecommand{\newblock}{\relax}
\providecommand{\bibinfo}[2]{#2}
\providecommand{\BIBentrySTDinterwordspacing}{\spaceskip=0pt\relax}
\providecommand{\BIBentryALTinterwordstretchfactor}{4}
\providecommand{\BIBentryALTinterwordspacing}{\spaceskip=\fontdimen2\font plus
\BIBentryALTinterwordstretchfactor\fontdimen3\font minus
  \fontdimen4\font\relax}
\providecommand{\BIBforeignlanguage}[2]{{%
\expandafter\ifx\csname l@#1\endcsname\relax
\typeout{** WARNING: IEEEtran.bst: No hyphenation pattern has been}%
\typeout{** loaded for the language `#1'. Using the pattern for}%
\typeout{** the default language instead.}%
\else
\language=\csname l@#1\endcsname
\fi
#2}}
\providecommand{\BIBdecl}{\relax}
\BIBdecl

\bibitem{Power2016}
D.~J. Power, ``Data science: supporting decision-making,'' \emph{J. Decis.
  Sys.}, vol.~25, no.~4, pp. 345--356, Apr. 2016.

\bibitem{Shah2016}
D.~Shah, ``Data science and statistics: Opportunities and challenges,''
  \emph{Technol. Rev.}, Sep. 2016.

\bibitem{SmithMRPCSSGLTST2008}
T.~Smith, N.~Maire, A.~Ross, M.~Penny, N.~Chitnis, A.~Schapira, A.~Studer,
  B.~Genton, C.~Lengeler, F.~Tediosi, D.~d. Savigny, and M.~Tanner, ``Towards a
  comprehensive simulation model of malaria epidemiology and control,''
  \emph{Parasitology}, vol. 135, no.~13, p. 1507–1516, Aug. 2008.

\bibitem{PKSMKFSAHiette2016}
J.~D. Piette, S.~L. Krein, D.~Striplin, N.~Marinec, R.~D. Kerns, K.~B. Farris,
  S.~Singh, L.~An, and A.~A. Heapy, ``Patient-centered pain care using
  artificial intelligence and mobile health tools: protocol for a randomized
  study funded by the us department of veterans affairs health services
  research and development program,'' \emph{JMIR Res. Protocols}, vol.~5,
  no.~2, 2016.

\bibitem{Nelson2007}
J.~Nelson, ``The operation of non-governmental organizations (ngos) in a world
  of corporate and other codes of conduct,'' \emph{Corporate Social
  Responsibility Initiative}, Mar. 2007.

\bibitem{CDC2005}
\BIBentryALTinterwordspacing
``{CDC/ATSDR} policy on releasing and sharing data,'' Sep. 2005. [Online].
  Available: \url{https://www.cdc.gov/maso/policy/releasingdata.pdf}
\BIBentrySTDinterwordspacing

\bibitem{VanPEGWHHB2014}
W.~G.~V. Panhuis, P.~Paul, C.~Emerson, J.~Grefenstette, R.~Wilder, A.~J.
  Herbst, D.~Heymann, and D.~S. Burke, ``A systematic review of barriers to
  data sharing in public health,'' \emph{BMC Public Health}, vol.~14, no.~1, p.
  1144, Feb. 2014.

\bibitem{CromanDEGJKMSSSSW2016}
K.~Croman, C.~Decker, I.~Eyal, A.~E. Gencer, A.~Juels, A.~Kosba, A.~Miller,
  P.~Saxena, E.~Shi, E.~G. Sirer, D.~Song, and R.~Wattenhofer, ``On scaling
  decentralized blockchains,'' in \emph{Financial Cryptography and Data
  Security}, ser. Lecture Notes in Computer Science, J.~Clark, S.~Meiklejohn,
  P.~Y.~A. Ryan, D.~Wallach, M.~Brenner, and K.~Rohloff, Eds.\hskip 1em plus
  0.5em minus 0.4em\relax Berlin: Springer, 2016, vol. 9604, pp. 106--125.

\bibitem{RemyBB2018}
S.~L. Remy, O.~Bent, and N.~Bore, ``Reshaping the use of digital tools to fight
  malaria,'' \emph{arXiv:1805.05418 [cs.CY]}, May 2018.

\bibitem{BentRRW2017}
O.~Bent, S.~L. Remy, S.~Roberts, and A.~Walcott{-}Bryant, ``Novel exploration
  techniques (nets) for malaria policy interventions,'' \emph{arXiv:1712.00428
  [cs.AI]}, Dec. 2017.

\bibitem{GundersonK2018}
O.~E. Gundersen and S.~Kjensmo, ``State of the art: Reproducibility in
  artificial intelligence,'' in \emph{Proc. 32nd AAAI Conf. Artif. Intell.},
  New Orleans, USA, Feb. 2018.

\bibitem{HendersonIBPPM2017}
P.~Henderson, R.~Islam, P.~Bachman, J.~Pineau, D.~Precup, and D.~Meger, ``Deep
  reinforcement learning that matters,'' \emph{arXiv:1709.06560v2 [cs.LG]},
  Nov. 2017.

\bibitem{SinghCN2018}
J.~Singh, J.~Cobbe, and C.~Norval, ``Decision provenance: Capturing data flow
  for accountable systems,'' \emph{arXiv:1804.05741 [cs.CY]}, Apr. 2018.

\bibitem{VermaCC2018}
D.~Verma, S.~Calo, and G.~Cirincione, ``Distributed ai and security issues in
  federated environments,'' in \emph{Proc. Workshop Program 19th Int. Conf.
  Distrib. Comput. Netw.}, ser. Workshops ICDCN '18, Jan. 2018.

\bibitem{Grier2011}
D.~A. Grier, ``Error identification and correction in human computation:
  Lessons from the wpa.'' in \emph{Human Computation}, 2011.

\bibitem{Androulaki2018}
E.~Androulaki, A.~Barger, V.~Bortnikov, C.~Cachin, K.~Christidis, A.~D. Caro,
  D.~Enyeart, C.~Ferris, G.~Laventman, Y.~Manevich, S.~Muralidharan, C.~Murthy,
  B.~Nguyen, M.~Sethi, G.~Singh, K.~Smith, A.~Sorniotti, C.~Stathakopoulou,
  M.~Vukoli\'{c}, S.~W. Cocco, and J.~Yellick, ``Hyperledger fabric: A
  distributed operating system for permissioned blockchains,'' in \emph{Proc.
  13th EuroSys Conf.}, ser. EuroSys '18, Apr. 2018, pp. 30:1--30:15.

\bibitem{TapscottT2016}
D.~Tapscott and A.~Tapscott, \emph{Blockchain Revolution: How the {T}echnology
  behind {B}itcoin is {C}hanging {M}oney, {B}usiness, and the {W}orld}.\hskip
  1em plus 0.5em minus 0.4em\relax New York: Penguin, 2016.

\bibitem{IansitiL2017}
M.~Iansiti and K.~R. Lakhani, ``The truth about blockchain,'' \emph{Harvard
  Bus. Rev.}, vol.~95, no.~1, pp. 118--127, Jan. 2017.

\bibitem{Vukolic2017}
M.~Vukoli\'{c}, ``Rethinking permissioned blockchains,'' in \emph{Proc. ACM
  Workshop Blockchain, Cryptocurrencies and Contracts}, ser. BCC '17, Apr.
  2017, pp. 3--7.

\bibitem{Tsai2018}
J.~Tsai, ``Transform blockchain into distributed parallel computing
  architecture for precision medicine,'' in \emph{2018 IEEE 38th Int. Conf.
  Distrib. Comput. Systems (ICDCS)}, Jul. 2018, pp. 1290--1299.

\bibitem{OzercanIAA2018}
H.~I. Ozercan, A.~M. Ileri, E.~Ayday, and C.~Alkan, ``Realizing the potential
  of blockchain technologies in genomics,'' \emph{Genome Research}, 2018.

\bibitem{FalconeST2001}
R.~Falcone, M.~Singh, and Y.-H. Tan, \emph{Trust in cyber-societies:
  integrating the human and artificial perspectives}.\hskip 1em plus 0.5em
  minus 0.4em\relax Springer Science \& Business Media, 2001, vol. 2246.

\bibitem{Marsh1994}
S.~P. Marsh, ``Formalising trust as a computational concept,'' Ph.D.
  dissertation, University of Stirling, 1994.

\bibitem{Dasgupta2000}
P.~Dasgupta, ``Trust as a commodity,'' \emph{Trust: Making and breaking
  cooperative relations}, vol.~4, pp. 49--72, 2000.

\bibitem{RamchurnHJ2004}
S.~D. Ramchurn, D.~Huynh, and N.~R. Jennings, ``Trust in multi-agent systems,''
  \emph{Knowl. Eng. Review}, vol.~19, no.~1, pp. 1--25, 2004.

\bibitem{GrangerJ1980}
C.~W. Granger and R.~Joyeux, ``An introduction to long-memory time series
  models and fractional differencing,'' \emph{J. Time Ser. Anal.}, vol.~1,
  no.~1, pp. 15--29, Jan. 1980.

\bibitem{GershoG2012}
A.~Gersho and R.~M. Gray, \emph{Vector Quantization and Signal
  Compression}.\hskip 1em plus 0.5em minus 0.4em\relax Springer Science \&
  Business Media, 2012, vol. 159.

\bibitem{ServettoVS1999}
S.~D. Servetto, V.~A. Vaishampayan, and N.~J.~A. Sloane, ``Multiple description
  lattice vector quantization,'' in \emph{Proceedings DCC'99 Data Compression
  Conference (Cat. No. PR00096)}, Mar. 1999, pp. 13--22.

\bibitem{EquitzC1991}
W.~H.~R. Equitz and T.~M. Cover, ``Successive refinement of information,''
  \emph{{IEEE} Trans. Inf. Theory}, vol.~37, no.~2, pp. 269--275, Mar. 1991.

\bibitem{MukherjeeM2002}
D.~Mukherjee and S.~K. Mitra, ``Successive refinement lattice vector
  quantization,'' \emph{{IEEE} Trans. Image Process.}, vol.~11, no.~12, pp.
  1337--1348, Dec. 2002.

\bibitem{LiuP2007}
Y.~Liu and W.~A. Pearlman, ``Multistage lattice vector quantization for
  hyperspectral image compression,'' in \emph{Conf. Rec. 41st Asilomar Conf.
  Signals, Syst. Comput.}, Nov. 2007, pp. 930--934.

\bibitem{MarshallO1960}
A.~W. Marshall and I.~Olkin, ``Multivariate {C}hebyshev inequalities,''
  \emph{Ann. Math. Stat.}, vol.~31, no.~4, pp. 1001--1014, Dec. 1960.

\bibitem{LecunBBH1998}
Y.~Lecun, L.~Bottou, Y.~Bengio, and P.~Haffner, ``Gradient-based learning
  applied to document recognition,'' \emph{Proc. IEEE}, vol.~86, no.~11, pp.
  2278--2324, Nov. 1998.

\end{thebibliography}
%%%%%%%%%%%%%%%%%%%%%%%%%%%%%%%%%%%%%%%%%%%%%%%%%%%%%%%%%%%%%%%%%%%%%%%%%%%%%%%%

\end{document}